%% file: ms_LC.tex
\documentclass[usegraphicx,usenatbib,natbib,useAMS]{mn2e}

\usepackage{times}     
\usepackage{aas_macros}
\usepackage{amssymb}   
\usepackage{amsmath}
\usepackage{multirow}

\def\halpha{\mbox{H$\alpha$}}
\def\hbeta{\mbox{H$\beta$}}

\def\lya{\mbox{Ly$\alpha$}}

\def\lesssim{\mathrel{\hbox{\rlap{\hbox{\lower4pt\hbox{$\sim$}}}\hbox{$<$}}}}
\def\gtrsim{\mathrel{\hbox{\rlap{\hbox{\lower4pt\hbox{$\sim$}}}\hbox{$>$}}}}

\DeclareRobustCommand{\ion}[2]{%
\relax\ifmmode
 \ifx\testbx\f@series
  {\mathbf{#1\,\mathsc{#2}}}\else
  {\mathrm{#1\,\mathsc{#2}}}\fi
 \else\textup{#1\,{\mdseries\textsc{#2}}}%
\fi}

\title[A break in the high-z Tully--Fisher
  relation]{A break in the high-redshift stellar mass Tully--Fisher
  relation \thanks{Based on data from the X-shooter GTO observations
    collected at the European Southern Observatory VLT/Kuyuen
    telescope, Paranal, Chile, under programme IDs: 084.B-0351(D),
    086.A-0674(A), 086.A-0674(B), 087.A-0432(A) and 087.A-0432(B). }}

\author[Christensen \& Hjorth]
       {Lise Christensen\thanks{Email: lise@dark-cosmology.dk} and
         Jens Hjorth\\
        $^1$ Dark Cosmology Centre, Niels Bohr Institute, University
         of Copenhagen, Juliane Maries Vej 30, 2100 Copenhagen,
         Denmark\\
}
\date{Accepted 2017 June 03. Received 2017 June 02; in original form 2016 January 17}
      
\pubyear{2013}

\begin{document}

\maketitle

\label{firstpage}

\begin{abstract}
   We investigate the stellar-mass Tully--Fisher relation (TFR)
   between the stellar mass and the integrated gas velocity
   dispersion, quantified by the kinematic estimator $S_{0.5}$
   measured from strong emission lines in spectra of galaxies at
   $0<z<5$. We combine luminosity-selected galaxies (`high-luminosity
   sample') with galaxies selected in other ways (`low-luminosity
   sample') to cover a range in stellar mass that spans almost five
   orders of magnitude: $7.0 \lesssim \log M_*/M_\odot \lesssim
   11.5$. We find that the logarithmic power-law slope and
   normalisation of the TFR are independent of redshift out to
   $z\sim3$.  The scatter in the TFR is $<0.5$ dex such that the gas
   velocity dispersion can be used as a proxy for the stellar mass of
   a galaxy independently of its redshift. At $z>3$ the scatter
   increases and the existence of a correlation is not obvious.  The
   high-luminosity sample exhibits a flatter slope of 1.5$\pm$0.2 at
   $z<3$ compared to the low-luminosity sample slope of 2.9$\pm$0.3,
   suggesting a turnover in the TFR.  The combined sample is well fit
   with a break in the TFR at a characteristic stellar mass scale of
   $M_*\sim10^{10}M_\odot$, with no significant evolution out to
   $z\sim3$.  We demonstrate that a break in the TFR with a steeper
   slope at the low-mass end is a natural consequence of galaxy models
   with a mass-dependent stellar to halo-mass ratio.

\end{abstract}


\section{Introduction}

Scaling relations for galaxies can provide insight into their
formation and evolution, with the caveat that sometimes the relations
depend strongly on selection effects. Disk galaxies are known to
follow the Tully--Fisher relation (TFR) \citep{tully77}, for which the
luminosity $L$ correlates with the maximum disk rotational velocity
$L\propto V^{3.5}$. The power-law slope depends somewhat on the
adopted photometric band.  While the traditional TFR studies of
rotational velocity versus luminosity include galaxies with regular
disk morphologies, investigations have also shown that both early- and
late type galaxies follow the same TFR in the local universe
\citep{trujillo-gomez11,cortese14}.  As the luminosity of a galaxy is
roughly proportional to its total stellar mass, there is also a
stellar-mass TFR.  Compared to the luminosity based TFR, local
galaxies have a steeper stellar-mass TFR (independent of photometric
band) with $M_* \propto V^{4.3}$ \citep{bell01}.

Several investigations have examined the evolution of the stellar-mass
TFR with redshift \citep[see][for a comprehensive
  review]{glazebrook13}, and although sample sizes are small at
higher redshifts, observations have revealed that the TFR slope and
normalisation is constant to $z\sim1.5$
\citep{milvang-jensen03,conselice05,flores06,miller11,miller12}. 
A small scatter in the TFR is obtained only when selecting
ordered disk galaxies without complex morphologies 
\citep{flores06,puech08}. Other studies report redshift evolution of the TFR
with a change in normalisation of 0.3--0.5 dex towards lower mass at a
given velocity between $z\sim0.6$ and $z\sim2$
\citep[][]{puech08,cresci09,vergani12}, and at $z=3$ a change in the
TFR normalisation downwards by 1 dex has been reported, albeit with a
large uncertainty  \citep{gnerucci11}.

Aside from the small samples sizes involved in studying the
high-redshift TFR, one caveat is that the dynamical range of galaxy
masses is limited since spectroscopic observations have primarily
pre-selected the brighter and more massive galaxies
\citep[e.g.][]{erb06a,foerster-schreiber09,contini12}.  This
necessitates assuming a constant TFR slope in determining the TFR
relation.  The issue of redshift evolution of the TFR (or lack thereof) 
therefore remains unsettled \citep{miller12}.

The two quantities that are primarily used to study galaxy disk
kinematics are the rotation velocity $V_{\mathrm{rot}}$ and the
velocity dispersion, $\sigma$. Velocity dispersions can be difficult
to derive in high-redshift quiescent galaxies, but it is easily
measured from emission-line widths in star-forming galaxies. At high
redshifts, galaxies have higher ratios of velocity dispersion to
rotation velocity \citep{genzel06,foerster-schreiber09}, revealing
complex dynamics and reflecting that disks are not yet settled.
Whereas high-spatial resolution adaptive optics observations coupled
with integral field spectroscopy can disentangle the rotation and
dispersion components, it is important to recall that such
investigations primarily target some of the most massive, and hence
luminous, galaxies at these redshifts. Obtaining the same level of
detailed kinematics in large samples of low-mass high-redshift
galaxies is not yet technically feasible.

To address these shortcomings, a variant of the classic TFR involving
a combination of the velocity dispersion and the rotation velocity
\citep{weiner06},
\begin{equation}
S_{K} = \sqrt{KV_{\mathrm{rot}}^2+\sigma^2},
\end{equation}
better represents the depths of the potential wells of galaxies that
are either dispersion or rotation dominated, or unsettled disks that
are common at higher redshifts.  Using this combination with $K=0.5$,
\citet{kassin07} found a smaller scatter compared to the classic TFR
for galaxies at $z\lesssim 1.2$, although \citet{miller11} found no
improvement by including a rotational component in the relation.
Using data from the SAMI- integral field data survey of 235 nearby
galaxies, \citet{cortese14} reported a tight $M_*$--$S_{0.5}$ relation
for both gas and stellar components in galaxies of any Hubble type.
Extending the analysis with literature data on massive galaxies at
higher redshifts, including a few up to $z=3.8$, \citet{kassin12} did
not find any indications of redshift evolution  of the TFR,
  although they did find that the relation between the rotation
  velocity and velocity dispersion does change with redshift.

 The classic TFR has a small scatter when choosing galaxies with
 well-ordered disks and smooth rotation curves. Including galaxies
 with a range of morphologies and luminosities cause a larger scatter,
 which may increase at higher redshifts as galaxies have increasing
 disordered rotations. The aim of this paper is to investigate the
 $M_*-S_{0.5}$ relation for a wide selection of galaxies reported in
 the literature. Rather that trying to build a complete sample of
 luminosity-selected galaxies that by far dominate the sample papers
 in the literature, we here explore if galaxies selected through
 alternative methods show a similar relation and scatter as
 luminosity-selected galaxy samples.  In addition to
 luminosity-selected samples we therefore include galaxies that are
 selected independently of their intrinsic luminosities and hence
 probe the lower-mass and lower-luminosity end of the
 distribution. Taking advantage of strong gravitational lensing, and
 deeper observations on faint galaxies, we study galaxies in a wide
 dynamical range of almost five orders of magnitude in stellar mass
 from $7\lesssim \log M_* \lesssim 12$, and extend the TFR to
 $z\sim5$.  In Section \ref{sect:data} we describe the data set, and
 explore the linear- and non-linear TFR in
 Sections~\ref{sect:scalings} and Section~\ref{sect:nonlin}. We
 briefly investigate relations for the star-formation rates in
 Section~\ref{sect:sfr}. Section~\ref{sect:discussion} discusses the
 findings and compares with other investigations, and
 Sect.~\ref{sect:conclusion} presents the summary.

\section{Data sets}
\label{sect:data}
As the stellar mass TFR has been well studied at $z\lesssim1$ using
luminosity selected samples
\citep{weiner06,kassin07,kassin12,cortese14}, this paper primarily
focuses on collecting higher redshift samples. We compile stellar
masses and integrated velocity dispersions from the literature, and,
when available, we also compile the inferred rotational velocities
($V_{\mathrm{rot}}$). Although the primary goal is to investigate
galaxies at $z>1$, we also include a local sample of 16 galaxies at
$z\sim0.2$ that are analogues to high-redshift Lyman break galaxies
\citep{goncalves10} in order to establish a low-redshift baseline for
comparison to the high-redshift galaxy samples.

With the advent of effective spectrographs on large telescopes, the
kinematics of hundreds of galaxies at $z\sim2$ have now been
studied. To date, by far the largest number of well-studied galaxies
at $z\sim1-3$ are relatively massive luminosity-selected ones
\citep{pettini01,erb06b,foerster-schreiber09,cresci09,law09,wright09,perez-montero09,lemoine10a,lemoine10b,gnerucci11,swinbank12,queyrel12,contini12,epinat12,lehnert13,kulas13,steidel14,wisnioski15}. Some
individual galaxies appear in several papers. We avoid including the
same galaxy twice. AO observations combined with integral field data
have been used to analyse spatially resolved kinematics of galaxies,
rather than a single measurement of their integrated velocity
dispersions \citep{foerster-schreiber09,gnerucci11,epinat12}.
Integral field spectra are vital to recover the maximum rotation
velocity of the galaxies, because slit-orientations may not follow
exactly the rotational axis of the galaxies.  Roughly half of the
galaxies in the luminosity-selected samples included in this paper are
observed with integral-field spectral data, where galaxy inclinations
can be measured, and the inferred rotational velocities include the
modelled inclinations.  Moreover, integral field spectra can reveal
variations in the measured velocity dispersion across the
galaxies. For resolved spectroscopic measurements we adopt the
reported integrated or average velocity dispersions. When only the
line FWHM are reported, we assume that the line has a Gaussian line
profile and $\sigma$ = FWHM/2.35.

To examine a large dynamical range in galaxy masses, we need data from
low-mass galaxies at higher redshifts. The main limitation is that in
order to resolve the emission lines and derive the intrinsic velocity
dispersions, which are of the order of a few tens of km~s$^{-1}$, a
resolving power of $R>5000-10000$ is needed. Low-mass, faint
high-redshift galaxies observed at intermediate resolution will have
spectra with lower signal-to-noise ratios, and the samples sizes are
consequently limited.  The galaxy luminosity function at $z>2$ has a
steep slope at the low-luminosity end, but these low-mass galaxies are
rarely targeted specifically for follow-up spectroscopy except in
special circumstances that use different selection methods.

Taking advantage of strong 
gravitational lensing by foreground clusters of galaxies is a way to 
mitigate the problem of low S/N ratio data, and several lensed galaxies have 
recently been observed at a sufficiently high spectral resolution to allow the
derivation of the velocity dispersion. In this paper we use data for
44 lensed galaxies at $1<z<5$ complied from the literature
\citep{teplitz00,siana08,swinbank09,hainline09,bian10,pettini10,christensen10,jones10,richard11,wuyts11,christensen12a,christensen12b,wuyts12,yuan12,jones13,belli13},
again avoiding duplications.

Other selection methods also probe preferentially the low-mass end of
the galaxy mass function. These include the host galaxies of explosive
events such as gamma-ray bursts (GRBs) or supernovae (SNe).  We
include 39 mostly low stellar-mass galaxies that hosted GRBs and which
have measured stellar masses\footnote{{\tt http://www.grbhosts.org/}}
\citep{savaglio09,perley13,kruhler15}. Although the total number of
well-investigated GRB hosts is more than 100, few are observed at
sufficiently high spectral resolution to allow a measurement of the
intrinsic velocity dispersion. For a few host galaxies, $\sigma$ or
the FWHM of emission lines are not reported, and we retrieved and
analysed ESO/X-shooter archive data in these cases. Original values
presented in various papers on GRB host galaxies are summarised in
Table~\ref{tab:grbhosts}.

Because of the limited sample sizes, we also include galaxies selected 
using other methods, such as two host galaxies of Type Ia SNe at 
$z\sim1.5$ \citep{frederiksen12a,frederiksen12b}, and a host galaxy
of a super-luminous SN \citep{leloudas15}, nine \lya\ emission
selected galaxies \citep{rhoads14}, three DLA galaxies
\citep{peroux11,fynbo13,krogager13},
and finally 14 extreme emission line galaxies \citep{maseda13}.

The rotational velocity component for these non-luminosity-selected
galaxies is rarely measured and reported along with the velocity
dispersion. One of the limiting factors is the spatial resolution
which prohibits the determination of the rotational velocity. 
Fortunately, galaxies at high redshifts are increasingly dominated by their 
intrinsic velocity dispersions \citep{genzel06,foerster-schreiber09}.
Moreover, for spatially unresolved galaxies, the one-dimensional spectra 
provide an integrated velocity dispersion which includes a rotational
component due to beam smearing. It can be shown that both dispersion 
dominated galaxies and galaxies with pure rotation, viewed edge on, have 
a line width of $S_{0.5}$ \citep[][]{rhoads14}. Therefore, $S_{0.5}$ is 
measured irrespectively of whether the galaxy is dominated by ordered or 
random motions.

The stellar masses reported in the literature and which we use in this
paper are based on conventional spectral energy distribution (SED)
fitting techniques, which find the best match among a set of galaxy
spectral template models to broad-band photometric data points.
Deriving stellar masses for galaxies above $z\sim3$ depend critically
on the availability of infra-red photometry without which stellar
masses might be underestimated. The luminosity-selected galaxy sample
is observed with Spitzer, while that is not the case for all other
selections. This will cause an increased scatter of the measurements
at the very highest redshifts. One of the dependent parameters in SED
fits is the stellar initial mass function (IMF) used in creating
galaxy spectral templates.  Authors typically either use a Salpeter
\citep{salpeter55} or a Chabrier IMF \citep{chabrier03}. When
necessary, we convert the reported total stellar masses to a Chabrier
IMF by dividing the mass inferred using a Salpeter IMF by a factor of
1.8. When reported, we also compile the star-formation rates (SFRs) of
the galaxies, again corrected to a Chabrier IMF.  Different
diagnostics are used to infer the galaxies SFRs, e.g. [OII], \halpha,
or UV luminosities. As these trace the SFRs on different time scales
but mostly agree within a factor of $\sim$2, the compiled values have
an increased scatter.  Additionally, some stellar masses and SFRs are
reported adopting different cosmologies, and we convert the reported
stellar masses to that that of a flat cosmology with
$H_0=70$~km~s$^{-1}$~Mpc$^{-1}$ and $\Omega_{\Lambda}=0.73$.

\begin{figure}
\begin{center}
\vspace*{6.4cm} \includegraphics[bb=561 69 78 677,clip, angle=-90,width=8.5cm]{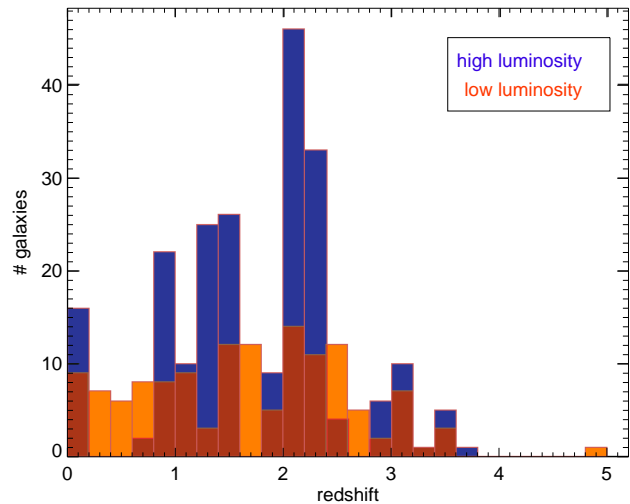}
\end{center}
\caption{Redshift distributions of the high- and low-luminosity
  samples. The high-luminosity sample consists of 214 luminosity
  selected galaxies.  The low-luminosity sample consists of 113
  galaxies selected using alternative methods, such as gravitationally
  lensed galaxies, GRB-, SNe-, DLA- host galaxies, Ly$\alpha$ emitters
  and extreme emission line galaxies.}
\label{fig:zdist}
\end{figure}

To summarise, the sample of galaxies can be split according to two distinct
selection methods: Galaxies that are selected via their luminosities
and colours, or via properties less directly related to their
brightness, namely lensed galaxies, DLA hosts, \lya\ emitters, GRB and
SN hosts, and extreme emission line galaxies. Broadly, the two samples
comprise high-luminosity galaxies and low-luminosity galaxies, respectively.
Although there is some overlap in luminosities between the two
samples, the terminology we adopt here is to refer to them as the
`high-luminosity' and `low-luminosity' galaxies. The combined sample
consists of 327 galaxies at redshifts $0<z<5$, 214 of which are in the
high-luminosity sample and 113 in the low-luminosity sample. The
redshift distributions illustrated in Fig.~\ref{fig:zdist} reveal
that luminosity selection preferentially targets galaxies at specific
redshift intervals where H$\alpha$ falls in the $H$ and $K$ bands,
causing conspicuous peaks around $z\sim 1.4$ and at $z\sim2.2$,
respectively. Conversely, the low-luminosity sample does not exhibit
any clear overdensities in redshift space.

We compare our compilation to $\sim$200,000 star-forming galaxies from
SDSS-DR9 tabulated in the MPA--JHU database\footnote{The SDSS
  catalogue data is obtained from
  \texttt{http://www.sdss3.org/dr9/algorithms/galaxy\char`_mpa\char`_jhu.php}}.
Emission-line velocity dispersions are determined from the integrated
spectrum within the 3 arcsec SDSS fibres \citep{thomas13}, and are
corrected for instrument resolution \citep{tremonti04}.  Rotational
velocities are not reported for SDSS galaxies, but since velocity
dispersions are measured within the central 3 arcsec fibre aperture,
beam smearing of any rotational component will also contribute to the
$S_{0.5}$ parameter.  Stellar velocity dispersions based on SDSS
  spectra are commonly used to measure dynamical masses. Emission-line
  velocity dispersions are less frequently used for the same purpose,
  but is has been shown that in star-forming SDSS galaxies the ratio
  between the two measurements is close to 1 \citep{chen08}.  We
select star-forming galaxies with well-detected emission lines and
emission-line ratios \citep[according to the BPT diagram
  in][]{baldwin81}, and with velocity dispersions derived from
forbidden emission lines. The reported stellar masses are based on a
Kroupa IMF and are therefore corrected down by a factor of 1.06 to
convert to a Chabrier IMF.  We choose only galaxies with
well-determined values of the velocity dispersion
($\sigma_{\mathrm{SDSS}}$) i.e., measurements with a signal-to-noise
ratio $>$3.

\section{Scaling relations for star forming galaxies}
\label{sect:scalings}
In this section we explore how selections of galaxy samples influence
the stellar mass TFR and its evolution with redshift.

\subsection{Stellar mass Tully--Fisher relation}
Using the combination of rotation and velocity dispersion in Equation
1 with $K=0.5$ for galaxies at $z\lesssim1.2$, \citet{kassin07} found
a smaller scatter than when examining the TFR based only on the
velocity dispersion (i.e., $K=0$ in Eq.\ 1).  We adopt $K=0.5$ for the
primarily high-mass, luminosity-selected galaxies that have a
measurement of $V_{\mathrm{rot}}$, while for the remaining galaxies
the reported velocity dispersions are representative for the $S_{0.5}$
parameter.  $V_{\mathrm{rot}}$ is reported for half of the galaxies in
the high-luminosity sample through measurements from IFU
data. However, the analysis and results reported below are consistent
if we investigate correlations with the velocity dispersion ($\sigma$)
alone.  Few of the galaxies have a much higher rotational component
compared to their velocity dispersions, and with median values of
$\sigma$= 82 km$^{-1}$ and $S_{0.5}$=111 km~s$^{-1}$ for the samples
with IFU data, the fits do not change. The intrinsic scatter of the
relations do however increase by 50\% for the samples at $z<2.2$.

\begin{figure*}
\begin{center}
\includegraphics[bb=60 360 1180 640,clip,  width=17.5cm]{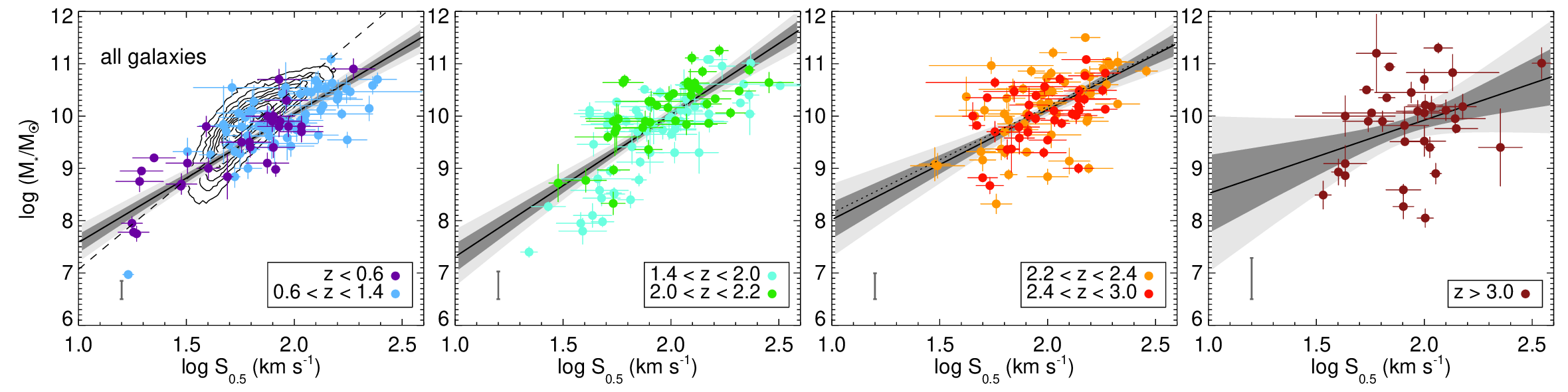}
\includegraphics[bb=60 360 1180 640,clip,  width=17.5cm]{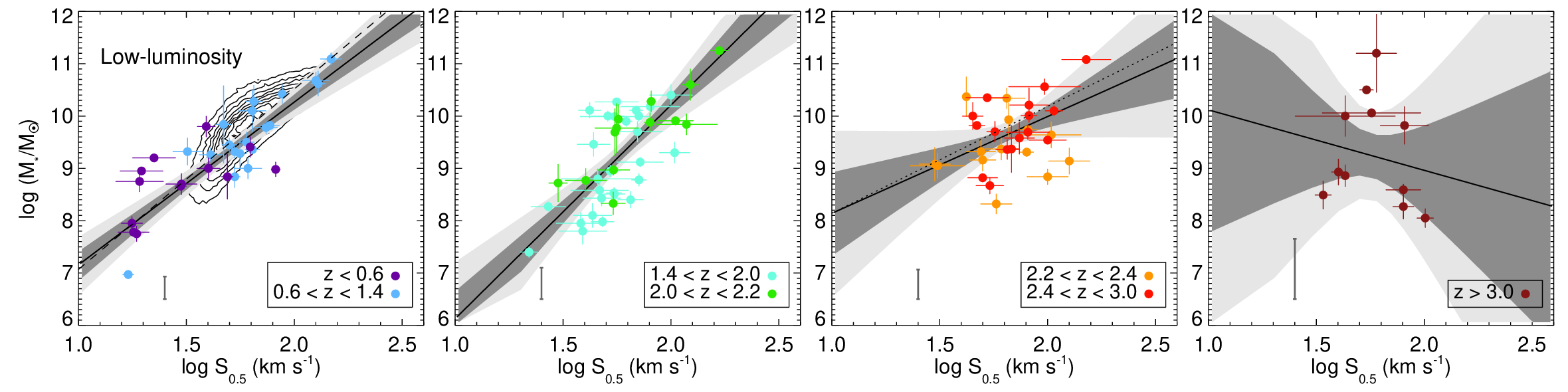}
\includegraphics[bb=60 360 1180 640,clip, width=17.5cm]{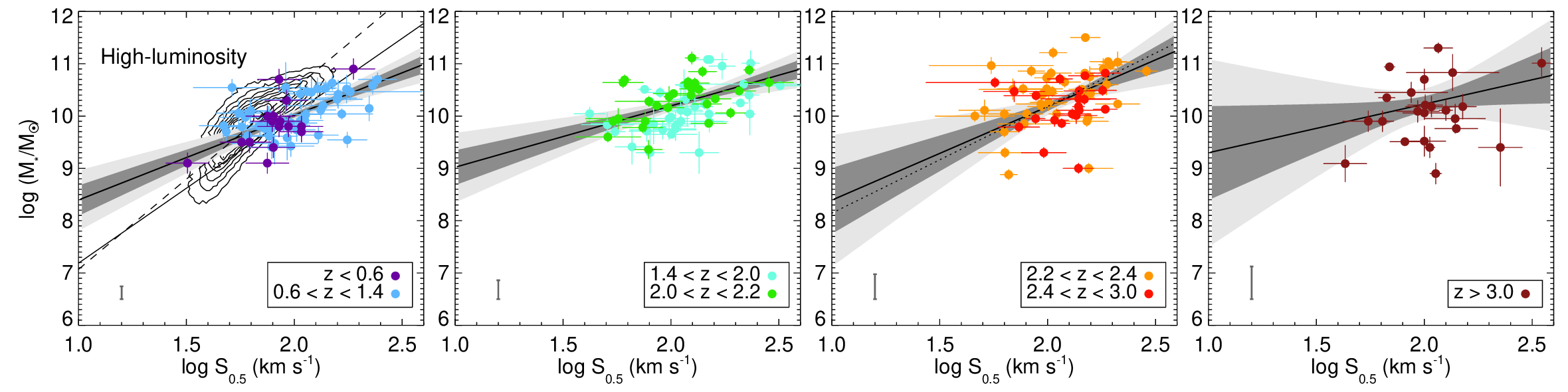}
\end{center}
\caption{Stellar mass TFR in different redshift bins. All stellar
  masses are rescaled to those obtained with a Chabrier IMF
  \citep{chabrier03}.  The top row of panels show all galaxies in this
  study, the middle row the low-luminosity galaxies, and the bottom
  row the high-luminosity galaxies. The contours in the first column
  represent star-forming SDSS galaxies. The dashed line in the first
  column shows the relation derived from luminosity-selected $z\sim1$
  galaxies \citep{kassin07}, and the dotted line in the third column
  refers to galaxies at $z>2$ \citep{barro14}. The linear fits are
  obtained with the {\sc linmix err} code \citep{kelly07}, and slopes
  and normalisations are reported in Table~\ref{tab:lin_params}. The
  grey shaded regions represent 1- and 2$\sigma$ confidence intervals
  respectively. Data points in the first three columns are colour
  coded in smaller redshift bins for visualisation purposes.  The bar
  in the lower left corner in each panel represents the intrinsic
  scatter of the relation, and explains why galaxies fall outside the
  shaded confidence regions.  }
\label{fig:msigma}
\end{figure*}

Figure~\ref{fig:msigma} shows the TFR for all the galaxies in our
compilation.  The sample up to $z\sim3$ has been split into three
redshift bins with roughly equal numbers of galaxies in each bin.  We
fit the TFR with a linear function (in log space) taking into account
the measurement uncertainties for the stellar masses and
velocities. Where these errors are not reported, we assume a
representative uncertainty of $\pm$0.2 dex in stellar mass (43 cases),
and $\pm$0.1 dex in log $S_{0.5}$ (33 cases). To determine the slope
and normalisation we use the {\sc linmix err} IDL code
\citep{kelly07}, which employs a Bayesian approach and a Marcov chain
Monte Carlo algorithm for the linear regression. It accounts for all
measurement errors and includes an intrinsic scatter term.  {\sc
  linmix err} assumes symmetric error bars so we symmetrised all
uncertainties (in log space).  We choose {\sc linmix~err} because it
is widely used and tested to examine for example the linear relation
between black hole masses and velocity dispersions, is more robust to
outliers.

Specifically, we fit a linear relation
\begin{equation}
\log M_* = A (\log S_{0.5} - 2.0)  +  B \pm \epsilon,
\label{eq:linfit}
\end{equation}
where $\epsilon$ represents an intrinsic scatter of the relation.  The
subtracted value 2.0 is chosen because the slope and intercepts are
strongly correlated, and it presents a value close to the median value
$\log~S_{0.5}=1.93$ of all galaxies.  A fit to all galaxies
irrespective of their redshift gives \(\log M_* = (2.40\pm0.15)\times
\log S_{0.5} + (5.25\pm0.29)\) with a scatter of $\epsilon=0.50$ dex
in log $M_*$. In comparison, \citet{kassin07} find a slope of
$2.94\pm0.38$, a normalisation of $5.56\pm0.82$, and a scatter of
$\epsilon=0.47$ dex in $M_*$ for all their galaxies at $0.1 < z <
1.2$.

Another linear chi-square minimising
routine like {\sc fitexy} \citep{numericalrecipes} which includes both
uncertainties of $S_{0.5}$ and $M_*$, with its modification to include
an intrinsic scatter, {\sc mpfitexy} \citep{williams10}, gives the
best fit \(\log M_* = (2.45\pm0.14)\times \log S_{0.5}+
(5.14\pm0.25)\) with a scatter of 0.49 dex in log $M_*$, consistent
with the results from {\sc linmix~err}.

\subsection{Redshift evolution}

To explore the evolution of the TFR with redshift, we divide the sample into
four redshift bins. Table~\ref{tab:lin_params} shows a compilation of
the best fit parameters and the number of galaxies in each redshift bin. 
Dividing the sample into high- and low-luminosity galaxies, the
best fit parameters are listed in
Table~\ref{tab:lin_params} and illustrated in the middle and bottom
rows in Figure~\ref{fig:msigma}.  At progressively higher redshifts,
the lower detection limit of both velocity dispersions and the stellar
masses increase.  This is a selection effect since the least massive
galaxies, which are also the least luminous ones, are not observed at 
sufficient spectral resolution.

\begin{figure}
\begin{center}
\includegraphics[bb=53 366 330 555, clip,width=8.5cm]{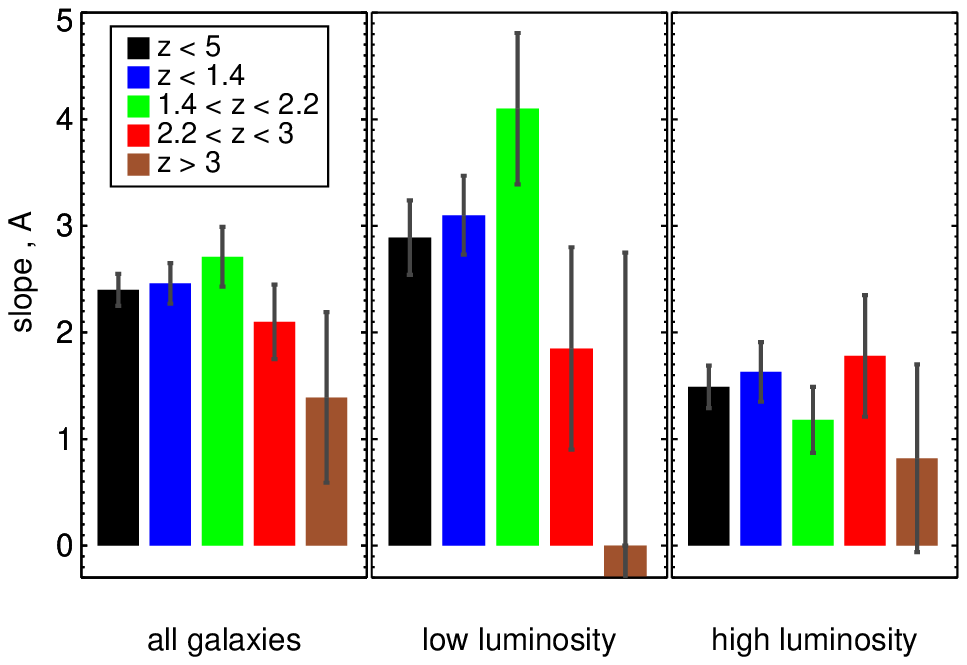}
\end{center}
\caption{Best fitting linear logarithmic slopes ($A$) in
  Table~\ref{tab:lin_params} are shown for the subsamples containing
  all galaxies, low- and high-luminosity galaxies. 
}
\label{fig:slope_plot}
\end{figure}

Within each sub-sample, there is no strong evidence for redshift
evolution at $z<3$. The best fit logarithmic slopes, $A$, are depicted
in Fig.~\ref{fig:slope_plot} (since the best fit slopes and
normalisations are highly correlated, an illustration of the
normalisation would be similar). However, there are differences in the
slopes of the two sub-samples. The low-luminosity galaxies have a
steeper slope in any redshift bin at $z<3$ with a slope of $2.9\pm0.3$
versus a slope of $1.5\pm0.2$ for the high-luminosity sample.  When
reversing the dependent and independent variables for the fits, we
find the same trend of a flatter slope (for the inverse relation) of
the low-luminosity relative to the high-luminosity sample at $z<2.2$,
a similar slope for the $2.2<z<3.0$ bin, and no relation for the
highest redshift bin. The steepening of the slope seen in
Fig.~\ref{fig:msigma} is therefore not caused by a paucity of low-mass
galaxies in the high-luminosity sample.

Previous investigations have analysed the change of the TFR
normalisation with redshift for small samples while fixing the slope
of the relation to a well-calibrated lower-redshift relation
\citep[e.g.][]{cresci09,gnerucci11,vergani12}. For consistency we
examine the result of keeping the slope fixed to that of the full
sample. We find no change in the best fit normalisation within
1-$\sigma$ uncertainties between all the redshift bins, again
suggesting that there is no redshift evolution.
 The TFR for galaxies at $z<1.2$ in \cite{kassin07} is consistent with
 that from our combined sample in the low-redshift interval in
 Table~\ref{tab:lin_params}.

\begin{table}
\setlength{\tabcolsep}{3pt}
\caption{Linear fits for the $M_*-S_{0.5}$ relation in
  Eq.~\ref{eq:linfit}. $\epsilon$ is the intrinsic scatter, $\rho$ is
  the linear correlation coefficient for the data, and $n$ is the
  number of galaxies in each bin.  The last column gives the
    covariance between the slope and intercepts. The top five rows
  fit all the galaxies, the middle five rows represent fits to
  low-luminosity galaxies, while the bottom five rows include the
  high-luminosity galaxies only.}
\begin{tabular}{crrrrrl}
\hline
\hline
Sample & $n$  & $A$ & $B$  & $\epsilon$ & $\rho$ & cov. \\
       &      &     &      &  (dex) \\
\hline %
All galaxies \\
$0.0<z<5.0$  &327& 2.40$\pm$0.15 & 10.05$\pm$0.03 & 0.50 & 0.73 & 0.001\\  
~~~~~~~~~$z<1.4$ &96 & 2.46$\pm$0.19 & 10.05$\pm$0.05 & 0.35 & 0.87 & 0.005\\
$1.4<z<2.2$  &106& 2.73$\pm$0.28 & 10.01$\pm$0.06 & 0.54 & 0.76 & 0.004\\
$2.2<z<3.0$  &91 & 2.11$\pm$0.35 & 10.13$\pm$0.06 & 0.49 & 0.64 & 0.0004\\
~~~~~~~~~$z>3.0$ &34 & 1.42$\pm$0.76 &  9.93$\pm$0.15 & 0.79 & 0.39 & 0.044\\
\hline
Low luminosity\\
$0.0<z<5.0$  &113& 2.87$\pm$0.35 & 10.06$\pm$0.11 & 0.61 & 0.69 & 0.03\\  
~~~~~~~~~$z<1.4$&32 & 3.10$\pm$0.37 & 10.26$\pm$0.15 & 0.43 & 0.90 & 0.04\\
$1.4<z<2.2$  &40 & 3.99$\pm$0.72 & 10.15$\pm$0.20 & 0.62 & 0.77 & 0.12\\
$2.2<z<3.0$  &30 & 1.86$\pm$0.96 &  9.99$\pm$0.21 & 0.57 & 0.48 & $-$0.08 \\
~~~~~~~~~$z>3.0$&11 & $-$1.29$\pm$3.00 & 8.95$\pm$0.83 & 1.15 & $-$0.24 & 2.14\\
\hline
High luminosity\\ 
$0.0<z<5.0$  &214& 1.49$\pm$0.21 & 10.13$\pm$0.03 & 0.41 & 0.53 & $-$0.002 \\  
~~~~~~~~~~$z<1.4$&64 & 1.63$\pm$0.30 & 10.02$\pm$0.05 & 0.25 & 0.76 & 0.0003\\
$1.4<z<2.2$  &66 & 1.18$\pm$0.31 & 10.19$\pm$0.05 & 0.36 & 0.52 & $-$0.087\\
$2.2<z<3.0$    &61 & 1.79$\pm$0.58 & 10.18$\pm$0.08 & 0.48 & 0.49 & $-$0.026\\
~~~~~~~~~~$z>3.0$&23 & 0.81$\pm$0.89 & 10.10$\pm$0.14 & 0.62 & 0.28 & $-$0.002\\
\hline
\end{tabular}
\label{tab:lin_params}
\end{table}

Expanding equation \ref{eq:linfit} to include a redshift term, we find
\begin{equation}
\begin{split}
\log M_* =& (2.37\pm0.14) [\log S_{0.5}-2.0] + (10.00\pm0.28) \\
          &  (-0.1\pm0.2)\log(1+z).
\end{split}
\label{eq:linfitz}
\end{equation}
The last term, \((-0.1\pm0.2)\log(1+z)\), minimises the scatter by a
negligible 0.002 dex and does not justify including a redshift
dependence. Fitting only galaxies that are not luminosity selected,
the best fit for the redshift evolution term is
\((-0.1\pm0.1)\log(1+z)\).  We conclude that there is no evidence for
significant redshift evolution of the TFR up to $z\sim3$.

A linear fit to the individual SDSS galaxies in Fig.~\ref{fig:msigma}
above a resolution limit of $\log \sigma_{\mathrm{SDSS}} > 1.8$ gives
$\log M_* = (2.34\pm0.09)\times (\log S_{0.5}-2.0) + (10.57\pm0.01)
\pm0.28$.  The slope is similar to that of the low-redshift sub-sample
including all the galaxies in Table~\ref{tab:lin_params}, within the
$1\sigma$ uncertainty range.  The difference in normalisation between
the SDSS and the sample containing all galaxies at $z<1.4$ is
$0.52\pm0.19$, suggesting a moderate redshift evolution. The change in
normalisation between the low-luminosity sample at $z<1.4$ and the
SDSS galaxies is smaller $0.31\pm0.15$. One reason for these different
normalisations is that the shape of the distribution of SDSS galaxies
does not appear to be linear, as we shall return to in
Sect.~\ref{sect:nonlin}.

Since $V_{\mathrm{rot}}$ is not measured for all the galaxies in the
sample, and only very few galaxies in the low-luminosity sample have
this value reported, it could affect the slope of the relation.
Instead of using $S_{0.5}$ we replace with the velocity dispersion,
$\sigma$ in all sub-samples and recompute the linear fits. Only in the
high-luminosity sub-sample at $z<1.4$ do the parameters change by more
than 1$\sigma$ uncertainties, giving a smaller slope compared to
Table~\ref{tab:lin_params}.  One data set that contributes to the
$z<1.4$ bin has a significantly higher $V_{\mathrm{rot}}$ compared to
$\sigma$ \citep{goncalves10}, and this drives the slope of the $M_* -
\sigma$ relation to a flatter value. Accordingly, we conclude that
using $S_{0.5}$ or $\sigma$ does not change the overall results
significantly. 

\section{A break in the $M_*-\sigma$ relation}
\label{sect:nonlin}
In the previous section we found that the slope of the TFR depends on
the luminosity range considered, with a more shallow slope for
luminous galaxies (Figs.~\ref{fig:msigma} and \ref{fig:slope_plot}).
We here quantify the existence of a break in the TFR and discuss a
possible interpretation.

\subsection{Asymptotic $M_*-\sigma$ relation}
\label{sect:asymp}

\begin{figure*}
\begin{center}
\begin{minipage}{0.5\textwidth}
\includegraphics[bb=67 360 685 785, width=\textwidth, clip]{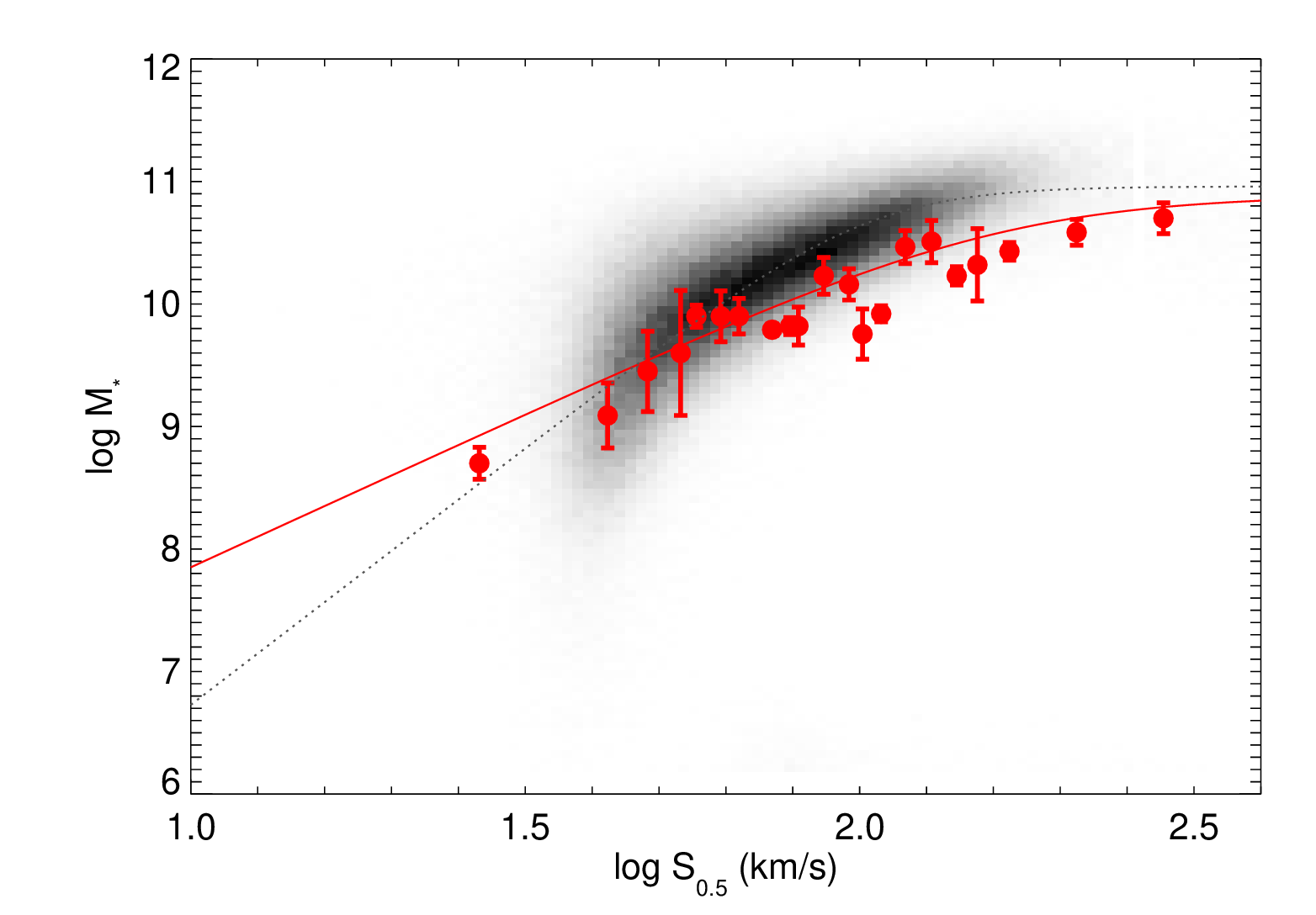}
\end{minipage}%
\begin{minipage}{0.5\textwidth}
\vspace*{2.92in}
\includegraphics[bb=550 111 83 710, angle=-90, width=0.98\textwidth, clip]{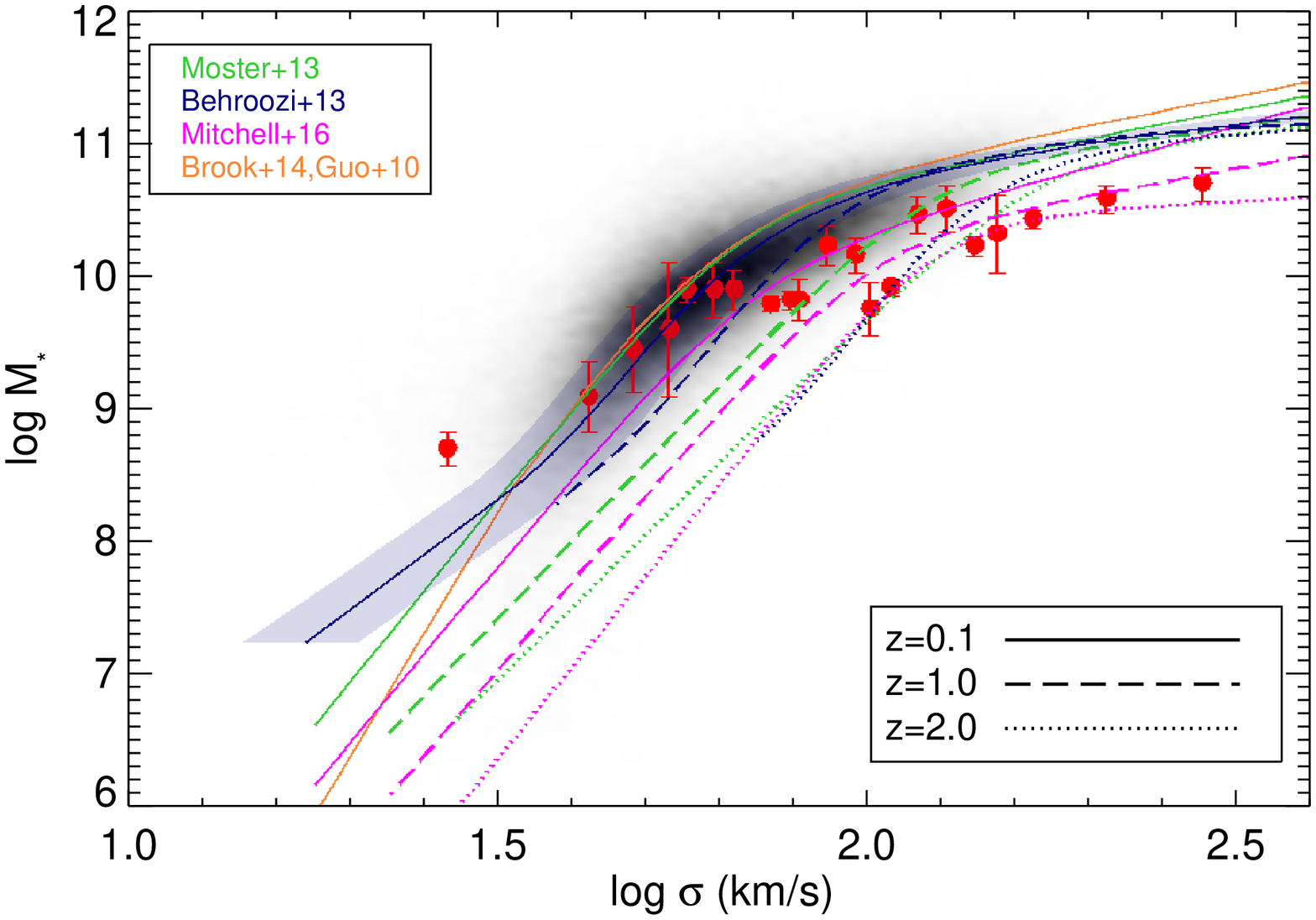}
\end{minipage}
\end{center}
\caption{A non-linear relation of the stellar-mass TFR. The galaxies
  in our compilation are separated into bins with equal number of
  galaxies in each bin (red circles). The error bars are calculated by
  bootstrapping the median masses in each bin.  The solid curve in the
  \textit{left panel} represents a fit to Eq.~\ref{eq:asym} using
  unbinned data. The grey scale distribution represent $\sim$200,000
  SDSS star-forming galaxies and the dotted curve is a fit to these.
  The fitting parameters suggests a similar asymptotic limit, while
  the turnover point for SDSS galaxies has a smaller velocity.
  \textit{Right panel: } The halo-mass to stellar mass fraction models
  \citep{guo10,moster13,behroozi13,brook14,mitchell16} combined with
  equations in Section~\ref{sect:halo_model} reproduce well the SDSS
  gas velocity dispersions at $z\sim0.1$. The shaded blue region is
  the 68\% confidence interval for the low redshift model from
  \citet{behroozi13}.  At higher redshifts, the models do not
  reproduce the observed data well, but a consistent fit can be
  obtained by changing the virial coefficient, $C$, to a higher
  value. In addition, a smaller stellar to total baryonic mass
  fraction at high redshift whould shift the data points downwards
  relative to the model predictions.}
\label{fig:msigma_bin}
\end{figure*}

To visualise the break in the relation we bin all galaxies in
$S_{0.5}$ (16 galaxies in each bin) and compute the median in each bin
(red filled circles in Fig.~\ref{fig:msigma_bin}). This procedure
implicitly assumes that there is no redshift evolution as suggested by
our findings in the previous section.  The uncertainty in the median
is calculated by bootstrapping. Fig.~\ref{fig:msigma_bin} suggests
that the turnover depends significantly on the lowest mass bin, which
is primarily composed of $1.4<z<2$ galaxies from the low-luminosity
sample (see Fig.~\ref{fig:msigma}, second column).  Twelve of the 16
galaxies in the lowest mass bin have been observed at intermediate- to
high spectral resolutions ($R>6000$), and their emission lines are
reported to be clearly resolved. If we exclude the other 4 galaxies
observed at lower spectral resolutions ($R\sim3000$), the lowest-mass
bin still reveals a break from a linear relation.

Assuming that this apparent break in the $M_*-S_{0.5}$ relation is not
a reflection of some unknown selection bias, we proceed to explore a
model composed of a low-luminosity power law and a high-luminosity
asymptotic (i.e., constant) limit,
\begin{equation}
\log M_* = \log M_{\mathrm{lim}} - \log\left[1 +
\left(\frac{S_{\mathrm{0.5,TO}}}{S_{0.5}}\right)^\gamma\right],
\label{eq:asym}
\end{equation}
where $M_{\mathrm{lim}}$ is the asymptotic limiting 
mass, $S_{\mathrm{0.5,TO}}$ is the turnover value, and $\gamma$ is the
power-law slope at the low $S_{0.5}$ end.
The fitting is done by $\chi^2$ minimisation in IDL (using the unbinned data), 
and the best fit is shown as the solid red curve in the left panel of
Fig.~\ref{fig:msigma_bin}, with best fit values reported in
Table~\ref{tab:params_assymp}. 
The low-mass end slope around $\gamma \sim$ 2.5 is similar to best fit
slopes for all galaxies in Section~\ref{sect:scalings}. $\gamma$
asymptotes to $A$ in the power-law (low-mass) limit.

 Introducing more parameters in a model to produce a better fit of the
 stellar mass TFR can be justified. Including the intrinsic scatter
 ($\epsilon$ in Equation 2) serves to make the reduced $\chi^2$ for
 the fit equal to 1. To evaluate if the asymptotic model provides a
 better fit than the linear model, we compute the Bayesian information
 criterion:
  \[\mathrm{BIC} = n \ln \Big(\sum^n_{i=1}(M_{*,i}-\hat M_i)^2/n
  \Big)+k \ln(n),\]
where $k$ is the number of parameters in the fit, $n$ is the sample
size, $M_*$ is the measured stellar mass, and $\hat M_i$ is the
expectation value from the model. For the binned sample, this gives
BIC = --55 for the linear fit, and BIC = --62 for the asymptotic fit,
and $\Delta(\mathrm{BIC})=7$ demonstrates a strong preference for the
latter model.

The grey scale area in Fig.~\ref{fig:msigma_bin} shows that SDSS
galaxies display a turnover at high velocity dispersions, where the
stellar mass approaches an asymptotic value. SDSS spectra have a
resolution of $R\sim2000$ corresponding to a velocity dispersion of
64~km~s$^{-1}$. As $\sigma_{\mathrm{SDSS}}$ is corrected for
instrument resolution, some line width measurements are very close to
the instrument resolution and should be treated with caution.
Emission lines with widths close to the instrumental resolution will
have higher relative uncertainties of their widths, and choosing only
very strong emission lines that satisfy the S/N criteria may skew the
values towards higher $\sigma$ at low stellar masses.  If we impose a
limit $\log \sigma_{\mathrm{SDSS}} > 1.6$, we find a turnover point at
$\log \sigma_{\mathrm{SDSS,TO}} = 2.0$ (see
Table~\ref{tab:params_assymp}), while a higher limit that represents
the SDSS velocity resolution of $\log \sigma_{\mathrm{SDSS}} > 1.8$
gives a turnover point at $\log \sigma_{\mathrm{SDSS,TO}} =
2.05\pm0.01$, a shallower low-mass slope $\gamma=3.20\pm0.04$ and a
similar limiting mass $\log
(M_{\mathrm{lim}}/$M$_{\odot})=11.09\pm0.01$.

Aperture effects in the SDSS spectra may also play a role for galaxies
with radii larger than the SDSS fiber size
\citep[e.g][]{brinchmann04,kewley05}. Imposing an additional criterion
that the 90 percentage Petrosian radius is smaller than 1.5 arcsec
leaves sample of just 235 SDSS galaxies. A fit to this small sample
gives a larger turnover point $\log \sigma_{\mathrm{SDSS,TO}} =
2.20\pm0.06$, a low-mass slope $\gamma=3.59\pm0.18$ and the same
limiting mass $\log (M_{\mathrm{lim}}/$M$_{\odot})=11.03\pm0.13$.  As
a turnover in the relation remains clear, we conclude that aperture
effects do not impose a strong effect on the determination of
$\sigma$. This is supported by the finding that emission line widths
and gas velocity dispersions do not display large changes with radius
in SDSS galaxies observed with integral field spectrographs
\citep[e.g.][]{gerssen12,garcia-lorenzo15}.

\begin{table}
\caption{Asymptotic parameter fits to Eq.~\ref{eq:asym}. }
\begin{tabular*}{0.5\textwidth}%
     {@{\extracolsep{\fill}}lllll}
\hline
\hline
Sample &  $\log M_{\mathrm{lim}}$ & $\log S_{0.5,TO}$ & $\gamma$  & scatter\\
       &  [log~M$_{\odot}$]    &  [km~s$^{-1}$] &          &  dex\\

\hline
SDSS         &  10.96$\pm$0.01 & 2.01$\pm$0.01 & 4.19$\pm$0.01 & 0.35\\

this sample  &  10.88$\pm$0.02 & 2.21$\pm$0.01 & 2.52$\pm$0.04 & 0.57\\

\hline
\end{tabular*}
\label{tab:params_assymp}
\end{table}

\subsection{Simple model with a non-linear relation}
\label{sect:halo_model}

The dynamics of a galaxy is governed by its total mass, and we can use
this information to predict the velocity dispersion for a galaxy with
a given mass. Simple equations lead to a scaling between the halo mass
and its virial velocity dispersion \(M_{\mathrm{halo}} \propto
\sigma_v^3\) \citep{posti14}. 

Models that match the dark-matter halo distribution with an observed
galaxy mass distribution predict that the stellar mass to dark matter
halo mass fraction depends on the halo mass. The stellar mass at the
maximum fraction is around $10^{12}$ M$_{\odot}$ at low redshifts
\citep{guo10}, and this stellar mass increases with redshift
\citep{behroozi13,moster13,mitchell16}. The standard definition of the
halo mass within a radius $r$ is
\begin{equation}
M_{\mathrm{halo}} = \frac{4}{3}\pi r^3 \Delta_c(z) \rho_c(z),
\label{eq:halo_rad}
\end{equation}
where the halo mass is typically defined within the virial radius of
the halo, $\Delta_c$ is the overdensity, and $\rho_c$, is the critical
energy density in a flat universe \(\rho_c(z) = \frac{3 H(z)^2}{8 \pi
  G}\) at redshift $z$.  The Hubble parameter evolves as \(H(z)=H_0
E(z) \) with $E(z)^2 = \Omega_{0,m}(1+z)^3+\Omega_{0,\Lambda}$ when
the radiation energy density can be neglected.  The overdensity can be
parametrised as
\begin{equation}
\Delta_c(z) = 18 \pi^2 +82[\Omega(z)-1] -39[\Omega(z) -1]^2
\end{equation}
\citep{bryan98,posti14},
where 
\begin{equation}
\Omega(z) = \Omega_{0,m}(1+z)^3/E(z)^2.
\end{equation}
In this parametrisation the overdensity at $z=0$ is $\Delta_c(0)=100$
and it increases with redshift.

The halo mass within a radius $r$ can be computed from the velocity
dispersion: \(M_{\mathrm{halo}}(<r) = C \sigma_v^2r/G\), where the
virial coefficient $C$ depends on the mass distribution within the
halo, including also velocity anisotropies and any assumptions of a
spherical or disk-like morphology of a galaxy. For an isothermal
sphere, $C=2$.  $C$ reflects the dominant contribution of the dark
matter to the halo mass, whereas in observations of galaxies it
primarily reflects galaxy properties measured within a typical
$\sim$10 kpc radius, where the baryons also contribute to the
mass. When deriving galaxy dynamical masses, values of $C=5$ for $r$
measuring the stellar half-light radius
\citep[e.g.][]{pettini01,shapley04}, or $C=3.4$ appropriate for
galaxies with disk-like morphologies \citep{erb06a} have been
chosen. Because $C$ depends on mass distribution and galaxy morphology
it will cause a scatter when deriving masses for a mixed galaxy sample
when set to a single value.

  Since we do not know the exact contribution of the halo mass within
  each galaxy and within the regions probed by the emission lines, we
  therefore take the approach that $C$ is not known, but is likely to
  be in the range of 1--10.

Combining the above equations yields
\begin{equation}
\label{eq:hscaling}
M_{\mathrm{halo}} = \frac{\sqrt{2}\sigma_v^3}{G}\frac{C^{3/2}}{H(z)\Delta_c(z)^{1/2}},
\end{equation}
i.e., $M_{\mathrm{halo}}$ scales as $\sigma_v^3$ and the
proportionality factor depends on the virial coefficient and redshift
through $\sqrt{C^3/\Delta_c(z)} / H(z)$.

Halo masses are converted into stellar masses using models of
stellar-to-halo mass fractions, and Equation~\ref{eq:hscaling}
produces a prediction for the relation between $M_*$ and
$\sigma_v$. The right panel in Fig.~\ref{fig:msigma_bin} shows the
binned data with overlayed halo-models
\citep{moster13,behroozi13,mitchell16} at redshifts $z$~=~0.1, 1 and
2. The model in \citet{brook14} that describes low-mass halos is
normalised to the higher-mass halo models in \citet{guo10}. All models
naturally have a break in the TFR due to the peak in the
stellar-to-halo mass fraction around $M_{\mathrm{halo}}=10^{12}$
M$_{\odot}$ \citep[see also Fig.~6 in][]{guo10}. Halo abundance
matching and computations of the rotational velocities also exhibit a
steepening below $\sim$100 km~s$^{-1}$ \citep{trujillo-gomez11}, and
reproduce the observed velocity-luminosity relation galaxies with
various morphologies including dwarfs, disk and giant elliptical
galaxies.

The exact point of the break location depends on $\Delta_c, C, H(z)$
and the chosen halo models which vary with redshifts. The low-redshift
models fit the SDSS data well for a choice of $C\approx2.5$. In
particular, the break in the relation and its normalisation are
reproduced. Since the horizontal location of the models depend on the
choice of $C$ and $\Delta_c$, other combinations could also provide a
reasonable agreement.  At higher redshifts however, some models fail
and predict too high stellar masses for a given velocity dispersion,
and the observed break is less pronounced than suggested by the
models. Decreasing the stellar-to-halo mass ratio by 0.3 dex would
improve the normalisation of the fits, and the break points of the
relations can shift to lower values of $\sigma$ for a larger value of
$C$. To reproduce the turnover point in the data, we determine that
$C\approx3-4$ and $C\approx6-7$ at $z=1$ and $z=2$, respectively. The
models illustrated in Fig.~\ref{fig:msigma_bin} all use a redshift
invariant $C=2.5$.

The average redshift of all the galaxies is $z=1.9$ and most are
selected by their star-formation properties in some form. Since they
have a higher SFR relative to galaxies in the local universe (see
Fig.~\ref{fig:sigma_sfr}), it is expected that they also have higher
gas-fractions, which could shift their location downwards in
Fig.~\ref{fig:msigma_bin}.  Deriving the stellar-to-halo mass ratio
based on semi-analytic models that include gas consumption and
conversion into stars results in a shift in normalisation with
redshift \citep{mitchell16}, and which reproduces better the
observations at the high-mass end. Low mass galaxies, on the other
hand, should also have a downward shift in normalisation because of a
higher gas fraction, but this shift is not seen in the data.

Although the simplistic approach to derive velocity dispersions based
on halo models do not reproduce perfectly the data at $z>1$, our
conclusion is that the existence of a break in the relation can
naturally be explained by a change in the stellar-to-dark-matter
fraction with galaxy mass. Furthermore, the models also predict a
shift of the turnover point towards higher velocity dispersions with
redshift consistent with the 0.2 dex change from local galaxies to
$z\sim2$ galaxies listed in Table~\ref{tab:params_assymp}.

\section{Star-formation rate versus $S_{0.5}$}
\label{sect:sfr}

\begin{figure*}
\begin{center}
\includegraphics[bb=60 360 1180 640,clip, width=17.5cm]{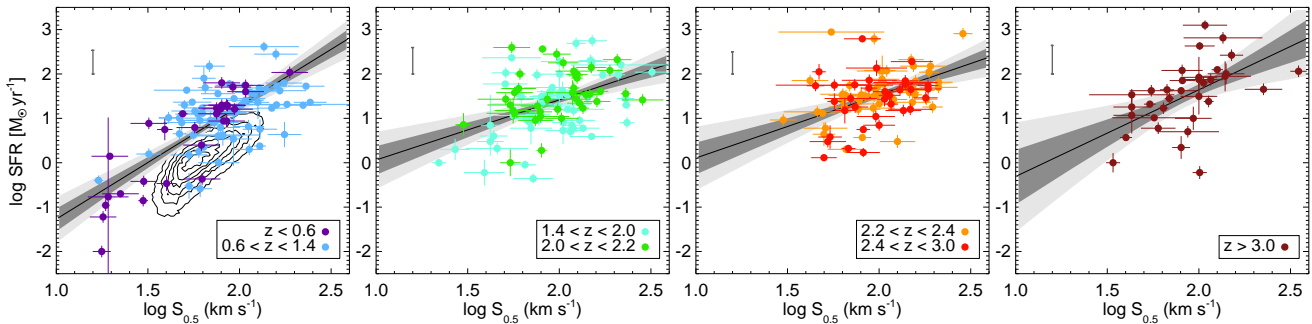}
\end{center}
\caption{SFR versus $S_{0.5}$ divided into redshift intervals.
  Uncertainties for the measured SFRs are typically smaller than the
  symbol sizes. The background contours in the left panel represent
  star-forming SDSS galaxies at $z\sim0$--0.4, which are clearly
  offset relative to the other galaxies.  The scatter in
    each panel is represented by the bar in the upper left corner.}
\label{fig:sigma_sfr}
\end{figure*}

\begin{table}
\caption{Linear relation fits for a SFR$-\sigma$ relation
  similar to Eq.~\ref{eq:linfit}.}
\begin{tabular}{llllll}
\hline
\hline
Sample & $n$  & $A'$ & $B'$  & $\epsilon'$ & $\rho$ \\
       &      &     &      &  (dex) \\
\hline
$0<z<5$      &308& 2.05$\pm$0.17 & 1.42$\pm$0.04 & 0.57 & 0.62\\
$z<1.4$      &95 & 2.55$\pm$0.27 & 1.29$\pm$0.07 & 0.53 & 0.77\\
$1.4<z<2.2$  &94 & 1.36$\pm$0.33 & 1.42$\pm$0.06 & 0.59 & 0.46\\
$2.2<z<3$    &87 & 1.43$\pm$0.35 & 1.52$\pm$0.06 & 0.50 & 0.48\\
$z>3$        &32 & 1.97$\pm$0.63 & 1.64$\pm$0.13 & 0.64 & 0.58\\
\hline
\end{tabular}
\label{tab:lin_params_sfr}
\end{table}

Other scaling relations involving the velocity dispersion have been
investigated. For example, \hbeta\ luminosities of local- and
high-redshift galaxies are found to be correlated with the widths of
the lines \citep{melnick00}. Spatially resolved observations of
galaxies at high redshifts also reveal that individual star-forming
clumps have velocity dispersions which scale with their luminosities
approximately as $L\propto \sigma^4$ \citep{wisnioski12}. Since
\hbeta\ luminosities scale linearly with the SFRs after correcting for
internal extinction, a correlation with the SFR must exist as well.

The integrated gas velocity dispersions could be affected by
star-formation and supernova feedback, in which case we would expect
to see a strong $\sigma$--SFR correlation. \citet{lehnert13} analysed
53 galaxies at $z=1$--3 observed with the VLT/Sinfoni integral field
spectrograph and concluded that $\sigma$ is driven by the SFR surface
density.  For 95 GRB hosts galaxies, i.e., non-luminosity selected
galaxies, \citet{kruhler15} investigated the $\sigma$-SFR relation and
found a strong correlation with a linear slope of $\sim$4. They also
reported marginal evidence (at the 1.5$\sigma$ level) for a redshift
evolution of the normalisation, assuming a fixed slope.

We fit all the galaxies with reported SFRs and $S_{0.5}$ to a linear
relation in Fig.~\ref{fig:sigma_sfr} with a best fit of: \( \log M_*
= (2.05\pm0.17) \times (\log S_{0.5} -2.0) + (1.42\pm0.04) \) with an
intrinsic scatter of 0.57 dex. As in Section~\ref{sect:scalings}, we
separate galaxies into redshift bins and report the fits in
Table~\ref{tab:lin_params_sfr}. In all redshift bins the intrinsic
scatter, $\epsilon$, is larger than in the $M_*-S_{0.5}$ fits. Using
the velocity dispersions, $\sigma$ instead of the $S_{0.5}$ parameter
does not change the scatter of the relation.

Compared to the results of \citet{kruhler15}, we derive a flatter
slope at all redshifts, but if we fix the slope to that obtained for
the full sample, the normalisation changes by 0.4 dex from the $z<1$
to the $z>3$ sub-samples. This change can be compared with the
$M_*$--SFR relation where a $\sim0.5$ dex increase in SFR from $z<1$
to $z>2$ for a given stellar mass has been inferred
\citep{whitaker14}.  Whereas the intercepts in
Table~\ref{tab:lin_params_sfr} are consistent within 1$\sigma$
uncertainties, the location of the 200,000 low-redshift star-forming
SDSS galaxies is clearly shifted to lower SFRs in
Fig.~\ref{fig:sigma_sfr}.  The change in normalisation of the
SFR-$S_{0.5}$ relation between SDSS galaxies and the high-luminosity-
and low-luminosity samples is 0.3 and 0.6 dex, respectively. Since the
$M_*$--$S_{0.5}$ relation appears constant with redshift out to
$z\sim3$, the reason for a varying SFR--$\sigma$ likely reflects the
observed mass-SFR scaling relation which is known to be redshift
dependent \citep{noeske07,daddi07}, as the redshift change in the
SFR-$S_{0.5}$ relation is similar to that seen in the mass-SFR scaling
relation \citep{whitaker14}.

\section{Discussion}
\label{sect:discussion}
The existence of a relation between the stellar mass and emission line
velocity dispersion parameter $S_{0.5}$ and its redshift evolution
has been frequently analysed in the literature. By combining various
galaxy samples and selection methods we have analysed the TFR covering
a large redshift span out to $z\sim5$. In this section we discuss how
various effects influence the results and compare with previous
studies.

\subsection{Sample selection}
Splitting up the galaxy samples into galaxies found in galaxy surveys
and flux-limited samples, and galaxies that are found via alternative
methods, we find that selection effects play an important role for the
interpretation of the TFR. In particular, galaxies selected via
alternative methods cover lower-mass
galaxies, and give rise to a steeper slope compared to luminosity
selected galaxies. 

The sample size and dynamical range of galaxy masses covered in
various surveys play a critical role. While we find that the TFR
relation extends at least out to $z\sim3$, a smaller sample of 22
extreme emission line galaxies at $1.4<z<2.3$ reveals no clear
relation \citep{maseda14}, while \citet{cortese14} find that low-mass
galaxies (log $M_*/\mathrm{M_{\odot}}<10$) have roughly the same
velocity dispersion of 20--30 km~s$^{-1}$.  A sample of $\sim$50
luminosity selected galaxies at $z\sim2$ suggested a correlation
\citep{foerster-schreiber09}, and \citet{erb06a} find a correlation
significance of 3.6$\sigma$ in a similar sample size.  In our larger
combined sample we see a clear trend that the TFR extends to lower
velocity dispersion in the $z<2$ samples.

\subsection{Evolution with redshift}
For the galaxies compiled in this study we find no significant changes
of the linear scaling between $\log M_*$ and $\log S_{0.5}$ with
redshift out to $z\sim3$. Figures \ref{fig:msigma} and
\ref{fig:slope_plot} illustrate that the slope remains constant to
within 1$\sigma$ uncertainties, while the scatter increases
significantly at $z>3$. The only difference in slopes arises from
sample selection as explained above.  We find the slope and
normalisation for the low-luminosity sample low-redshift bin to be
consistent with that determined for $z\sim1$ galaxies
\citep{kassin07}. The (linear) fit to SDSS galaxies suggests a shift
in the normalisation of $0.31\pm0.15$ dex from $z\sim0$ to the
low-luminosity sample at $z<1.4$, suggesting a minor redshift
evolution, although with the large uncertainties, it is also
consistent with being constant to within a $2\sigma$ level.

Previous investigations that extend the TFR to higher redshifts find a
shift of 1 dex at $z\sim3$ \citep{gnerucci11}, and \citet{cresci09}
find that the normalisation changes downwards by 0.4 dex for a sample
of 18 galaxies at $z\sim2$ compared to the local relation. Other
studies find that the $S_{0.5}$ relation shows no evolution with
redshift at $z<1.2$ \citep{kassin07}, and while \citet{kassin12} find
that it appears to be constant out to $z>3$, they suggest that the
reason is that the selection is biased towards higher mass galaxies.

Numerical simulations from EAGLE \citep{schaye15} and semi-analytical
models \citep{dutton11} were used to investigate the redshift
evolution of the classic TFR \citep{tiley16}.  These models suggest a
redshift evolution with a shift of about --0.2 dex from $z=0$ to
$z=1.5$. This shift is consistent with the normalisation change from
SDSS to our lowest redshift (low-luminosity) bin.  The models suggest
a further shift of --0.3 dex from $z=1.5$ to $z=3$, while the
normalisations for this redshift interval listed in
Table~\ref{tab:lin_params} imply a change of $+0.12\pm0.08$ for all
galaxies, or $-0.16\pm0.29$ for the low-redshift sample. Given the
large uncertainties and the scatter of the data points, we cannot rule
out any models based on our measurements.

When comparing the slopes and normalisations with previous studies, it
is important to bear in mind that we include all measurement
uncertainties and explore the relation over a larger dynamical range
in stellar mass and redshift interval. The choice of method for
fitting a straight line to the relation can also lead to different
results \citep[see][]{hogg10}. Choosing instead a simple linear
regression or a bisector fit which does not include either individual
measurement uncertainties or an intrinsic scatter term, gives steeper
slopes for all redshifts and all subsamples, and is consistent with
those previously reported ($A\sim3.5$), i.e., a slope that large
resembles the classical TFR or the local Faber--Jackson relations
slope of 4 \citep{faber76}.

\subsection{Scatter of the relations}  
Fitting linear relations (in log space), we find an intrinsic scatter
of all the subsamples in the range 0.2--0.6 dex with a larger scatter
in the low-luminosity sample compared to the high-luminosity
sample. In the $z>3$ bin the scatter is so high that the existence of
a relation is not evident.

The scatter we derive is similar to a scatter of 0.47 dex for galaxies
at $0.1<z<1.2$ \citep{kassin07}, 0.45 dex reported in
\citep{vergani12}, and $\sim$0.3 dex in $M_*$ for a local galaxy
sample \citep{cortese14}. A smaller scatter of 0.14 dex in
$\Delta\log\sigma$, equivalent to 0.29 dex in log $M_*$, has been
found for compact, massive star-forming galaxies at $z\sim2.3$
\citep{barro14}, with a selection that only covers a range of 1 dex in
stellar mass. The corresponding slope of 2.04 is similar to what we
find in the same redshift bin $2.2<z<3$ for the combined sample.

The rather large scatter we find is a natural consequence of not
restricting our sample to galaxies that are kinematically or
morphologically similar.  Studies of galaxies at $z=0.6-1.2$ have
revealed a roughly three-way split between rotating disks,
merger-dominated galaxies and dispersion-dominated ones. By selecting
only rotation dominated disks, the TFR scatter decreases significantly
\citep{puech08,vergani12}.

\subsection{Intepretations of a break in the TFR}
        
We find that the $M_*$--$S_{0.5}$ slope depends on the luminosity
range considered suggesting a non-linear relation (in log space). A
similar non-linear relation is also seen in low-redshift star-forming
SDSS galaxies. Fitting the data to a phenomenological model consisting
of a low-luminosity power-law saturating at a high-mass asymptotic
limit (Sect.~\ref{sect:asymp}) we find that the asymptotic mass limit
of local SDSS galaxies is very similar to the high-redshift sample,
with a turnover value of $\log S_{\mathrm{0.5,TO}}\sim2.0-2.2$
depending on the chosen resolution limit and potential aperture
losses. When studying the stellar mass TFR, \citet{cortese14} also
find a non-linear behaviour, and note that the $S_{0.5}$ relation
becomes steeper below $M_*=10^{10} M_{\odot}$.

Several effects could give rise to a break in the relation. Velocity
dispersion measurements are frequently used to compute the total
dynamical mass of early-type galaxies, and since both quiescent and
star-forming galaxies at low- and high redshifts follow a single TFR
\citep{cortese14,barro14}, velocity dispersions measured from emission
lines also trace the gravitational potential of a galaxy. However, in
gas-rich late-type galaxies, the gaseous component may contribute with
a significant amount of mass, and therefore the total baryonic mass
including stars and gas should be measured.

\citet{mcgaugh00} find a turnover in the $M_*-S_{0.5}$ relation with a
steeper low-mass slope, but also determine an increasingly large
gaseous component in low-mass galaxies. Without including the
additional gas-mass the TFR becomes steeper in the low-mass end.
Another example of a change in slope comes from \lya\ emitting
galaxies at $2<z<3$, which contain a larger gas-to-stellar mass
fraction relative to other galaxy types \citep{rhoads14}. These
galaxies lie systematically below the stellar-mass TFR in
\citet{kassin07}, but their dynamical masses show better agreement
with the relation. \citet{zaritsky14} find a turnover at a rotational
velocity at $\log v_c\approx2$ for low-redshift galaxies in the
classical TFR, and argue that the break can be corrected for when
including gas masses in the baryonic TFR. Specifically, they find that
including a constant fraction of halo-mass to baryonic mass of 7\%
will give rise to a linear TFR. If the galaxies investigated here were
to follow this constant halo-baryon fraction, the amount of missing
gas-mass in the lowest mass galaxies ($M_*\approx 10^{8-9}$
M$_{\odot}$), would be a factor of 20--10. While the amount of gas
present in these star-forming galaxies may be significant compared to
their stellar masses, such a high factor is larger than inferred for
low-mass galaxies (M$_*<10^{10}$ M$_{\odot}$) at $z\sim1$, which have
gas masses of a factor of $\sim$2 larger than their stellar masses
\citep{stott16}. While some galaxies, particularly at $z>3$, lie
significantly below the $z\sim1$ stellar-mass TFR, most galaxies
exhibit a spread around the relation with objects both above and below
the relation. Hence, an increased gas fraction in high redshift
galaxies cannot be the only explanation of a turnover.

Another effect that could change the slope primarily at the high-mass
end is the turbulent motion related to star-formation activity either
from in- or outflows, shocks or merger activities. Turbulent motions
increase the velocity dispersion and give rise to a flatter
relation. As star-formation activity increases by a factor of $\sim$10
from $z=0$ to $z>1$, this would imply a stronger effect in a higher
redshift sample.  However, there are indications that star-formation
activity does not significantly alter the location of galaxies in the
TFR. Galaxies of any Hubble types appear to follow the same
mass-$S_{0.5}$ relation \citep{cortese14} and compact quiescent
massive galaxies appear to follow the same relation as their
star-forming counterparts \citep{barro14}. In our samples, the slope
of the $S_{0.5}$-SFR relation becomes flatter with redshift which
could be interpreted as a dynamical effect in the high-mass end. As a
very high SFR influences a low-mass galaxy more, the specific SFR
(i.e. SFR/$M_*$) better probes the impact of the star formation
activity on the velocity dispersions. In our sample there are no clear
correlations between the specific SFR and $S_{0.5}$ for any of the
subsamples in any of the redshift bins.

 High-redshift galaxies are increasingly dispersion dominated
 \citep[e.g.][]{foerster-schreiber09}, and galaxies exhibit more
 ordered rotation with decreasing redshifts \citep{kassin12}. At $z<1$
 \citet{kassin12} find a threshold at $\log M_*=10.4$ where galaxies
 transition from low-mass disordered to more ordered kinematics. At
 any redshift, the most massive star-forming galaxies are the ones
 that have rotationally supported disks, and this transition
 progresses to lower masses with cosmic time.  Including a high
 rotational velocity component in addition to the measured velocity
 dispersion in the calculation of $S_{0.5}$ causes galaxies to deviate
 from a steep $\sigma-M_*$ relation, and will give rise to a
 flattening of the $S_{0.5}-M_*$ relation in the high-mass end. The
 down-sizing of galaxies with increasing rotational support with
 decreasing redshift implies a change in the turnover point to lower
 stellar masses and lower velocity dispersions. This is further
 suggested by the observed change in the turnover point between the
 SDSS and the higher redshift sample.

To summarise, the likely explanations for a turnover in the TFR, is
either a higher gas-mass fraction in low-mass galaxies, or a higher
dark matter to stellar mass fraction. Based on the arguments above,
and that the halo-models appear to broadly describe the observations
in section~\ref{sect:halo_model}, it seems that the dark-matter
contribution to the break is a valid interpretation, while a vertical
shift in Fig.~\ref{fig:msigma_bin} can be explained by a change in the
gas fraction.  Future observations of total gas masses in high
redshift galaxies are needed for a conclusion of the dominant effect.

\section{Summary}
\label{sect:conclusion}
By examining a variety of galaxy samples out to $z\sim5$, covering a
wide dynamical range in stellar masses, we find a redshift invariant
TFR between stellar mass and the kinematic estimator $S_{0.5}$ that
combines velocity dispersion and galaxy rotations derived from
emission lines. While there is no strong evidence for evolution out to
$z\sim3$, we find that sample definition and fitting methods play a
crucial role in the determined slopes and normalisations of the fits
to the TFR. This conclusion also holds when we extend the study to
low-redshift star-forming galaxies in the SDSS.  Between $z=0$ and
  $z=2$, the SFR--$S_{0.5}$ relation displays a redshift evolution
similar to the $M_*$--SFR relation, also known as the main sequence
for star-forming galaxies, suggesting that these redshift changes
arise from the same underlying physical evolution mechanism.

We find that the stellar mass TFR is not linear and has a steeper
slope at the low-mass end.  The reason for a turnover can be caused by
a several physical effects. A large fraction of baryons in
star-forming dwarf galaxies is in gaseous form, and adding this
component to the stellar mass is found to create a linear relation
over the whole dynamical range. One would expect a linear relation in
log space as the global velocity dispersion traces the dynamical mass
of the galaxy. In this study we have not investigated dynamical masses
because the galaxies radii are not measured for the majority of the
sample. Instead, we rely on the stellar mass alone which may not be
representative for the dynamics in the galaxy.

A larger rotation causes an increase in $S_{0.5}$ for more massive
galaxies and at lower redshifts when galaxy disks settle into more
ordered rotations. Since galaxies have more ordered rotation with
increased stellar mass and increased cosmic time, this would imply a
shift of the turnover towards lower values at lower redshifts.

A higher fraction of dark matter in low-mass galaxies also causes a
turnover of the relation and results in a steeper slope at the
low-mass end.  The break we infer from the observations broadly agree
with halo ocupation distribution models, which are derived by matching
modelled halo masses with observed galaxy stellar masses, and not the
galaxies' total baryonic masses \citep{guo10,behroozi13,moster13}.  We
find a good agreement at low redshifts, but at $z\sim2$, the
observations suggest a less pronounced break compared to model
predictions. At the high mass end, the observed galaxies have a lower
stellar mass than predicted, which can be explained by a higher gas
fraction and consequently a lower stellar mass fraction. Models that
include gas consumption and conversion into stars are better able to
reproduce the normalisation of the TFR.  At the low-stellar mass end,
the observations appear to have a too high stellar mass compared to
the models. However, the slope at the low-mass end, depends on
relatively few (only 20 galaxies with $\log S_{0.5} < 1.6$) low-mass,
high redshift galaxies. Further studies to explore the low-mass end of
high-redshift galaxy dynamics are needed to securely constrain the
low-mass end slope of the TFR relation.

\section*{Acknowledgements}
LC acknowledges support from an YDUN grant DFF 4090-00079.  This
research has made use of the GHostS database ({\tt www.grbhosts.org}),
which is partly funded by Spitzer/NASA grant RSA Agreement
No. 1287913.  We thank James Rhoads, Anja von der Linden, Arianna Di
Cintio and Chris Brook for useful discussions. We also thank the
referees for insightful comments to the manuscript.


\bibliographystyle{apj}
\bibliography{ms_LC}

\appendix
\input{grbtable}

\end{document}

%% file: grbtable.tex
\renewcommand{\thetable}{A\arabic{table}}

\begin{table}
\begin{tabular*}{0.5\textwidth}%
     {@{\extracolsep{\fill}} lllllll}

\hline
\hline
GRB name &  $z$ & log $M^*$ $^a$ & $\sigma$ & SFR  & ref \\
         &      & [$M_{\odot}$] & [km s$^{-1}$] & [M$_{\odot}$~yr$^{-1}$] & \\ 
\hline
GRB980425&   0.0085& 9.20$\pm$0.07&   22.4$\pm$0.2  & 0.2         &[1]\\
GRB000210&   0.8460& 9.31$\pm$0.08&   64$\pm$5      & 2.28        &[25]\\	
GRB000418&   1.1181& 9.26$\pm$0.14&   41$\pm$5      & 10.4        &[25]\\
GRB000911&   1.0578& 9.32$\pm$0.26&   32$\pm$6      & 1.57        &[25]\\
GRB011211&   2.1433& 9.77$\pm$0.47&   37.5$\pm$15.5 & 4.9         &[26]\\
GRB021004&   2.3299& 9.31$\pm$0.08&   68.8$\pm$11.0 & 2.28	  &[26]\\
GRB031203&   0.1055& 8.84$\pm$0.43&   49.0$\pm$3.0  & 12.7         &[2]\\
GRB030329&   0.1680& 7.75$\pm$0.15&   18.6$\pm$0.9  & 0.2          &[3]\\
GRB050416&   0.6540& 9.84$\pm$0.74&   47$\pm$4  & 1.5$\pm$0.2  &[4,24]\\
GRB050709&   0.1606& 8.66$\pm$0.07&   29.8$\pm$1.8  & 0.14         &[21,26]\\
GRB050915A&  2.5272  &10.56$\pm$0.15& 96.8$\pm$37.1 & 136$\pm$55   &[12,20]\\
GRB051022&   0.8061  &10.42$\pm$0.18& 88$\pm$5      & 60           &[24]\\  
GRB060218&   0.0335& 7.78$\pm$0.08&   18.0$\pm$3.0  & 0.06         &[5]\\
GRB060505&   0.0889& 9.41$\pm$0.01&   62.5$\pm$1.5  & 0.43         &[6]\\ 
GRB060614&   0.1250& 7.95$\pm$0.13&   17.7$\pm$2.0  & 0.01         &[7,26]\\
GRB060719&   1.5318&10.11$\pm$0.15&   42$\pm$5      & 7$\pm$11     &[12,20,24]\\
GRB060814&   1.9229& 9.99$\pm$0.05&   54.9$\pm$20.3 & 232$\pm$37   &[12,20]\\
GRB061126&   1.1588&10.62$\pm$0.24&  129.0$\pm$10.0 & 2.38         &[8]\\
GRB070306&   1.4959&10.4$\pm$0.2  &  133.0$\pm$10.0 & 34$\pm$29    &[9,10]\\
GRB070802&   2.4538& 9.7$\pm$0.2  &   63.6$\pm$29.4 & 10$\pm$18    &[4,10]\\
GRB071021&   2.4513&11.08$\pm$0.03&  151.5$\pm$39.7 & 190$\pm$23   &[12,20]\\
GRB080207&   2.0858&11.25$\pm$0.11&  167.5$\pm$16.5 & 40.7$\pm$1.6 &[4,11]\\
GRB080605&   1.6403& 9.90$\pm$0.10&   62.0$\pm$4.0  & 31.0$\pm$10  &[4,10]\\
GRB080805&   1.505 & 9.7$\pm$0.2  &   70.0$\pm$18.5 & 6$\pm$5      & [10,12]\\
GRB081109&   0.9787& 9.82$\pm$0.09&   77.2$\pm$7.8  & 13$\pm$4     & [10]\\
GRB090323&   3.5832& 11.2$\pm$0.75&   60.$\pm$13    &  6	   & [24] \\ 
GRB090407&   1.4474&10.18$\pm$0.15&   80.2$\pm$25.2 & 28.1$\pm$12.2& [10,12]\\
GRB090426&   2.609& 10.35$\pm$0.06&   52.5$\pm$12   &     3        & [13,14]\\
GRB090926B&  1.2425& 10.1$\pm$0.4&    63.9$\pm$9.4  &  80$\pm$50   & [10]\\
GRB091127&   0.4903& 8.70$\pm$0.20&   60.0$\pm$10.0 &  0.22        & [15]\\
GRB100418&   0.6239& 9.28$\pm$0.28&   56$\pm$4      &  4.2$\pm$0.9 & [24]\\
GRB100621A&  0.5420& 8.98$\pm$0.14&   82$\pm$4      &  8.7$\pm$0.8 & [10,24]\\
GRB110918A&  0.984 &10.68$\pm$0.16&   126$\pm$18    &  23$\pm$20   & [16]\\
GRB120119 &  1.7280 & 9.30        &   104$\pm$17    &  43$\pm$19   & [24]\\
GRB120422A&  0.2826& 8.95$\pm$0.04&   19.6$\pm$0.2  &  0.3$\pm$0.3 & [22]\\
GRB120624B&  2.1974&10.6$\pm$0.3  &  123.0$\pm$11.0 &  91$\pm$6.   & [17]\\	
GRB130427 &  0.3399& 9.0$\pm$0.2  &  40.0$\pm$5.0   &  0.34$\pm$0.13& [18,24]\\
GRB130603B&  0.3565& 9.8$\pm$0.2 &    39.1$\pm$3.8  &  5.6$\pm$3.1 & [19]\\
GRB140506A&  0.8893&  9.0   &     61$\pm$9	&     0.26$\pm$0.11 & [23,24]\\
\hline
\end{tabular*}
\caption{GRB host data. $^a$ Stellar masses are adopted from the GHostS compilation  in \citet{savaglio09}.  References.-- [ 1] \citet{christensen08},
  [2] \citet{guseva11}, [3] \citet{thoene07}, [4] \citet{hjorth12},
  [5] \citet{wiersema07}, [6] \citet{thoene08}, Th{\"o}ne et al. in
  prep. [7] \citet{hjorth05}, [8] \citet{perley08}, [9] \citet{jaunsen08},
  [10] \citet{kruhler11}, [11] \citet{svensson12}, [12]
  \citet{kruhler12}, [13] \citet{thoene11}, [14] \citet{levesque10},
  [15] \citet{vergani11}, [16] \citet{elliot13}, [17]
  \citet{deugarte13}, [18] \citet{xu13},
  [19] \citet{deugarte14}, [20] \citet{perley13}, [21]
  \citet{hjorth05}, [22] \citet{schulze14}, [23] \citet{fynbo14}, [24]
  \citet{kruhler15}, [25] T. Kr\"uhler, priv. comm.,
  [26] ESO xshooter archive, this paper.  }
\label{tab:grbhosts}
\end{table}